\documentclass[letterpaper, 11pt]{article}

\usepackage{graphicx}
\usepackage{pdfpages}
\usepackage{color}
\usepackage{framed}
\usepackage{hyperref}
\usepackage{url}
\usepackage[font=footnotesize,labelfont=bf]{caption}
\usepackage{float}
\usepackage{listings}
\usepackage{amsmath}
\usepackage{subfig}

\usepackage{algorithm}
\usepackage[noend]{algpseudocode}
\usepackage[margin=1.25in]{geometry}

\usepackage{tikz}
\usetikzlibrary{arrows,automata}

\graphicspath{ {images/} }
\title{\LARGE Plasma Cash: Towards more efficient Plasma constructions}
\author{
                Georgios Konstantopoulos\\
                \footnotesize\href{mailto:me@gakonst.com}
                        {\nolinkurl{me@gakonst.com}}
        }   

        \date{\today\\\small WORKING DRAFT }

\ifdefined \theorem
\relax
\else
\newtheorem{theorem}{Theorem}

\fi

\begin{document}

\maketitle

\begin{abstract}
    Plasma is a framework for scalable off-chain computation.
    We describe and evaluate Plasma Cash, an improved Plasma construction 
    which leverages non-fungible tokens and Sparse Merkle Trees to reduce the
    data storage and bandwith requirements for users. 
    We analyze the cryptoeconomic exit and challenge mechanisms used to keep user funds secured, even when the Plasma Cash chain's
    consensus algorithm is compromised. A reference implementation is provided
    for evaluation. Finally, we briefly discuss further improvements that
    can be made to the Plasma Cash protocol such as arbitrary denomination
    payments, less user data checking, fast and optimistic exits.
\end{abstract}

\tableofcontents

\section{Introduction}

\textit{Plasma} is a scalability solution to increase the 
throughput a smart contract enabled blockchain by allowing 
interoperable transfer of assets between blockchains. Assets are deposited 
from the sending blockchain (\textit{mainchain}), to
the receiving blockchain (\textit{plasmachain}).
Contrary to two-way pegged sidechains \cite{sidechains}, funds stored in
\textit{plasmachains} can safely be withdrawn to the \textit{mainchain} through a
dispute process where the mainchain is responsible for the safety of the coins, often referred to as an \textit{exit}. Safety is maintained even if the
plasmachain's consensus mechanism is compromised, as long as the
mainchain's consensus mechanism remains secure. It should be noted that
Plasma improves throughput\footnote{The number of transactions that can be
finalized every N seconds} rather than latency\footnote{The amount of time
before a transaction is confirmed}.

Plasma Cash leverages non-fungible tokens\footnote{A non-fungible token (NFT) is a unique token. Only one copy can exist for a specific NFT.} by making the funds deposited in the plasmachain independent and unique \cite{plasma_cash}. This results in a simpler settlement protocol for withdrawing funds back to the mainchain, by requiring less data checking on their coins, compared to the technique mentioned in the original paper \cite{plasma}. 
Discussions around this technique have been found across  the web, in ad-hoc 
discussions and videoconferences \cite{implementers_call}. To date there exists no 
complete resource on the security and scalability guarantees of Plasma Cash.

\subsection{Contributions} \label{contribs}

\paragraph{Protocol Analysis} We present a full analysis of the Plasma Cash
technique, decomposing it to its components, explaining parts of the
protocol and its functionalities in detail. We additionally explain the various
extensions which can be added, along with the benefits and
tradeoffs they give.

\paragraph{Analysis of Attacks} We discuss the currently known attacks on
Plasma and how a Plasma contract should be designed to mitigate these attacks. 
With proper mechanism design and the attachment of `cryptoeconomic' bonds on
actions that may result in faults we realize a system that is secure
against adversaries, even if the adversaries have full control of the plasmachain's consensus mechanism.

\paragraph{Implementation} A reference implementation of Plasma Cash is
provided at \url{github.com/loomnetwork/plasma-cash}. It provides support for
Ether and ERC20 tokens
\footnote{https://github.com/ethereum/EIPs/blob/master/EIPS/eip-20.md}, 
as well as ERC721 Non-Fungible tokens
\footnote{https://github.com/ethereum/EIPs/blob/master/EIPS/eip-721.md}. 
The Plasma contract is token agnostic and extending it to support more types of
token standards is a trivial process.

\subsection{Related Work}

In this section we will briefly describe work that is being done towards scalability via other techniques, each with its own tradeoffs. 
There are two approaches to scalability. Protocol level scalability\footnote{E.g. sharding and bigger blocks}, often referred to as
Layer 1, and scaling through computation via Layer 2. Plasma is considered a Layer 2 scalability solution.

\paragraph{Sidechains} A sidechain is a blockchain which has its own independently secured consensus algorithm, and 
is pegged to another blockchain \cite{sidechains}. Value can be transferred 
    from one blockchain to another by relaying SPV proofs\footnote{Simple Payment Verification: Proofs that allow clients to verify the inclusion of a transaction in a block by verifying the block headers instead of downloading the whole block}. Verifying 
    that blockchain A's event has occured on blockchain B requires SPV proofs which grow linearly with the number of blocks\footnote{Recent research has shown that the proof growth can be reduced to be logarithmic \cite{nipopows}}. It should be noted that users can only withdraw their funds back to the mainchain if the sidechain's consensus is secure and does not censor transactions or produce invalid state changes, meaning that the consensus mechanism is the sidechain's `custodian'. Using a sidechain for scalability is thus limited by the security and decentralization tradeoffs introduced by increasing capacity and throughput.

\paragraph{Payment Channel Networks} A payment channel is a mechanism which
enables two parties to trade value in a rapid and feeless manner by exchanging signed messages which represent the latest state of the channel. The state can be settled on the underlying blockchain at any point in time. If a party attempts to settle an earlier state, the other party can challenge/dispute within a bound time period and penalize the fraudulent party. An on-chain transaction is required to initialize the channel. Channels can be unidirectional, bidirectional and can be linked together to create Payment Channel Networks, which allow two parties to trade with each other, even if they do not have a channel with each other, by routing payments through intermediaries. Payment channels feature instant finality since a payment can be considered complete the moment signed attestations about it are exchanged from the participating parties. In addition to the on-chain funding transaction required during the initialization of a channel, Payment Channel Networks (cf. Lightning Network, Raiden Network) require a capital lockup equal to the value being
transferred, as well as potential fees paid to intermediaries for routing payments. 

\paragraph{Generalized State Channels} Extending the idea of payment channels, generalized state channels tackle the problem of performing arbitrary computation off-chain, with the ability to settle on-chain, e.g play a battleship game off-chain, with the ability to settle a dispute about the general state of the game on-chain \cite{battleship, counterfactual}. 

%

\section{Plasma Fundamentals}
\begin{figure}[H]
    \makebox[\linewidth]{
        \scalebox{0.75}{
            \includegraphics[width=\linewidth]{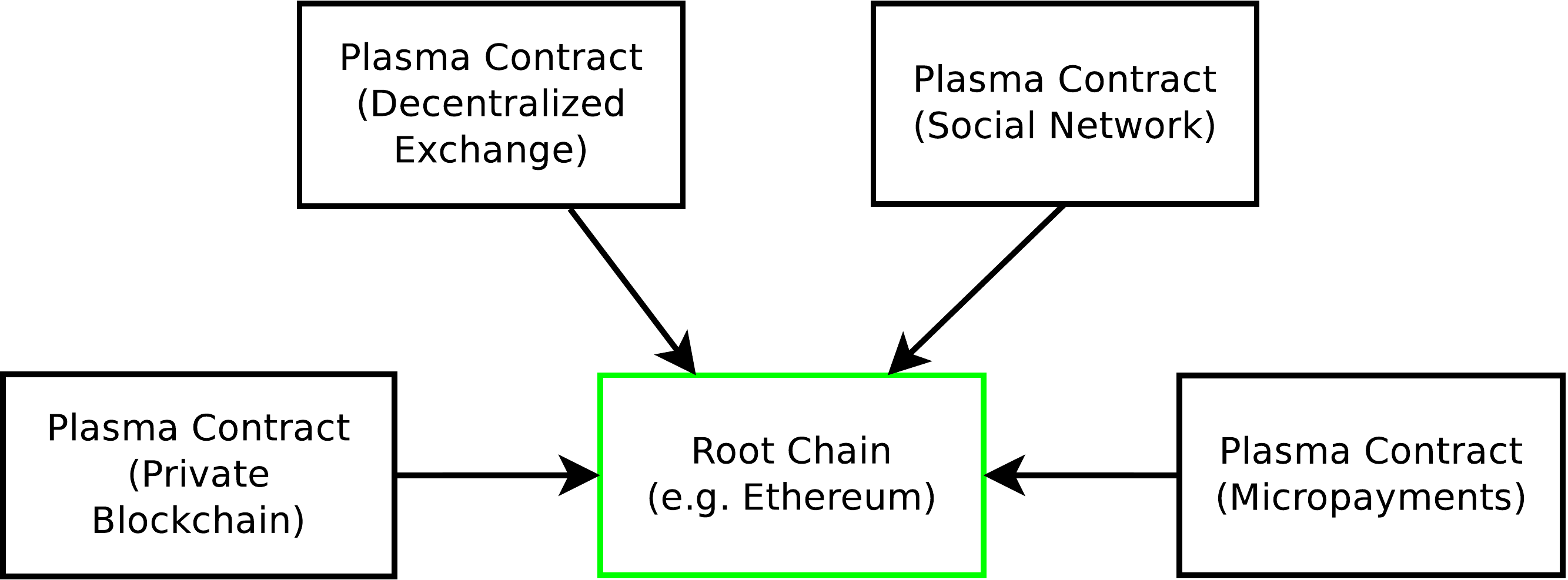}
        }
    }
    \caption{
        Multiple blockchains connected to a `mainchain' \cite{plasma}.
    }
    \label{fig:many-chains}
\end{figure}

\paragraph{`Classic' Plasma} \label{ch2:classic_plasma}
The initial vision of Plasma describes a mechanism which enables connecting
blockchains to a base blockchain often referred to as the `rootchain' or the
`mainchain' \cite{plasma}. The mainchain acts as the final arbitrator in
the case of disputes as shown in Figure \ref{fig:many-chains}. Trees of plasmachains forming a hierarchy of blockchains similar to the court system would also be possible such that a dispute can be escalated all the way to the supreme court (the mainchain)

A plasmachain that utilizes a potentially centralized consensus algorithm (e.g. Proof of Authority) derives its security from the mainchain's consensus algorithm. This is achieved by publishing a merklized\footnote{The transaction set of each block gets inserted in a Merkle Tree, and the merkle root of that tree gets published to a smart contract.} state root of each \textit{plasma block} to the mainchain. The transaction model used is UTXO based, similar to Bitcoin\footnote{https://en.bitcoin.it/wiki/Transaction}.

\paragraph{Exits and challenges} An \textit{exit} is the mechanism by which a user expresses
their intent to withdraw a UTXO from the plasmachain back to their account on
the mainchain. After submitting the exit of a UTXO, the user needs to wait for
a \textit{challenge period} to pass, during which other users can \textit{challenge}
their exit, by providing proof that it is invalid\footnote{This is a common technique also used in payment channels, whereby invalid attestations can be challenged and the fraudulent user gets penalized if the challenge is successful.}. 

A challenge is an attestation provided by another user, proving that the UTXO
that is being exited is invalid or has been spent. Challenges can either be
non-interactive, where the exit is instantly cancelled, or interactive where
they require a response to the challenge before the \textit{challenge period} is over. 
After the challenge period has passed, if there is no outstanding challenge, 
an exit can be finalized, giving its owner the right to withdraw the amount 
specified by the UTXO back to their mainchain account. 

Exits and interactive challenges require users to attach a bond, as collateral. If a user's exit gets challenged and the challenge is valid, the collateral gets slashed and is sent to the user who reported the fraud. This incentivizes users to act honestly, and creates a crowdsourced system of watchers who are rewarded for reporting fraudulent activity.

\section{Plasma Cash}
Plasma Cash is a plasma construction with much less user data checking \cite{plasma_cash}. It
utilizes Non-Fungible-Tokens (NFTs) to reduce the
user checking requirements to only the NFTs that they own\footnote{Plasma-MVP requires users to be constantly monitoring the plasmachain for fraudulent state transitions, while Plasma Cash requires that users only watch the mainchain about fraudulent exits of coins that they own}. The system's security relies on users fully authenticating a coin's history before accepting it as a payment by utilizing Sparse Merkle Trees, which allow the efficient verification of inclusion and non-inclusion of a transaction in a block, as explained in the next subsection.

\subsection{Sparse Merkle Trees} 

A Merkle Tree is a data structure which allows to succinctly commit to a dataset and prove the inclusion of a part of the committed dataset in $O(log_2(N))$ steps instead of $O(N)$, where $N$ is the number of elements in the dataset, via \textit{Merkle Proofs}. The committed value is called a \textit{Merkle Root}. A Sparse Merkle Tree (SMTs) \cite{sparse_merkle_trees} is an ordered merkle tree, where each element of the dataset is placed at the leaf with the index corresponding the element's index in the dataset. If an element of the dataset was not included in the Merkle Root, its leaf is set to a special default value. 

\begin{figure}[ht!]
        \centering
        \subfloat[Merkle proof of inclusion for coin A]{\includegraphics[width=.2\textwidth]{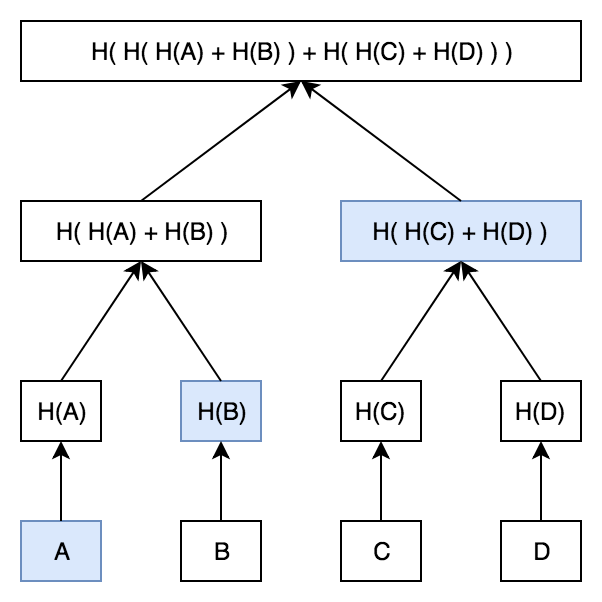}}%
        \qquad
        \subfloat[Merkle proof of non-inclusion for coin C]{\includegraphics[width=.2\textwidth]{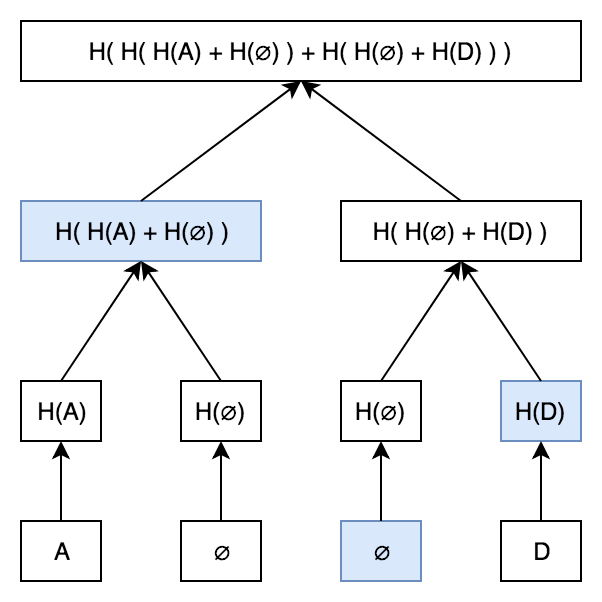}}%
    \caption{Merkle Proofs in a Sparse Merkle Tree \cite{smt}}
    \label{fig:smt}
\end{figure}

The inclusion of a transaction spending a coin in a block can be efficiently proven through a Merkle Proof. The same method can be used to prove the non-inclusion of a spend of a coin in a block. A coin can only be spent once per block because only 1 transaction at its slot can ever exist. If a coin was not spent, the leaf is set to the hash of 0. A visual representation of the above is given in Figure \ref{fig:smt}.

Further optimizations can be done to the verifier
as suggested in \cite{smt_compact_proofs} by precomputing the default values of
the SMT and by introducing a bitfield in the proof which acts as a
switch between choosing the next 32 bytes during the verification from the
proof or from the SMT's default hashes. A reasonable estimation for a block
containing 2378 transactions results in proof sizes being 320 bytes, compared
to normal proofs which would be 2048 (64 * 32) bytes, for a SMT of size 64.

\subsection{Periodic Merkle Commitments}
All Plasma designs rely on a plasmachain operator that commits the
Merkle Root of each generated block to the mainchain. If a Proof of Stake system is used, publishing a block must be accompanied by a number of validator signatures, exceeding a pre-agreed threshold. 
Whenever a Plasma Block's root is committed to the rootchain, all valid
transactions in that given block can be considered finalized upon availability of the related witness data\footnote{By witness data we refer to the merkle proof of inclusion of that coin in the specified block} for the inclusion of these transactions. Since only the
block root is committed to the mainchain instead of all the included
transactions, plasma can bundle together any number of transactions, the only limit to the number of included transactions being the size of the plasma block. The minimum finality time that
can be achieved is the block time of the rootchain (15 seconds in Ethereum).
Given that every block must be published to the mainchain, the operational costs for this process can become large. Operators can be expected to trade bigger finality time for less maintenance costs by committing block roots less often.

\subsection{Non Fungible Tokens and Depositing a Coin}

When value gets deposited in the plasma smart contract, a unique id and metadata key-value pair gets generated for that token and is saved in the contract's storage. The id is the unique serial number of the coin, and is what makes the deposited coin non-fungible. As a result, depositing 5 ETH two times creates two coins which have their own unique transaction history and are independent from each other. The unique id can also be thought of as a serial number, compared to fiat cash money.

After the coin gets saved in the smart contract, a block containing only the deposit transaction is appended to the plasmachain\footnote{Deposit blocks including only one transaction is an optimization, https://ethresear.ch/t/one-plasma-cash-block-per-deposit-why/2674/}.

%
\subsection{Transferring a Coin and Verifying its History}
\label{verify_coin_history}
Each coin in Plasma Cash has its own unique coin history. A coin receiver  must
verify that the coin they are receiving has a valid history in order to accept 
it. A coin with invalid history is counterfeit
and cannot be withdrawn safely. In order to validate a coin's
history, a set of merkle proofs of inclusion
and non-inclusion for the coin since its initial deposit must be sent to the receiver. The receiver can then
proceed to verify that there were no invalid spends of the coin in the coin's
history. This is done by verifying that the proofs of inclusion and non-inclusion are valid
against the plasma block merkle roots that were published to the mainchain.

This imposes a heavy storage and bandwith burden on senders that 
want to transfer a coin. Specifically, the proofs required to send a coin are 
$O(t * log_{2}(N))$ where $t$ is the number of blocks since a coin's deposit and $N$ is the number
of coins the Plasma Cash chain supports.

In a real world scenario where a buyer wants to buy a product from a vendor the
following is expected to happen, in a non-fraudulent case:
\begin{enumerate}
    \item Buyer broadcasts transaction giving ownership of their coin to the
        seller
    \item Transaction gets included in a block and witness data about its inclusion is made available
    \item Buyer verifies that the transaction was included in the block
    \item Buyer sends the proofs of inclusion and non-inclusion to the vendor.
    \item Vendor verifies the history of the coin along with the correct inclusion of the coin's transaction in the block.
    \item Vendor gives the product to buyer
\end{enumerate}

A transaction is a tuple: \texttt{Tx(slot, parentBlock, newOwner, prevOwnerSignature)}. A transaction that was included in a block is a combination of the previous tuple and a merkle proof for the block: \texttt{IncludedTx(tx, blkNumber, proof)}.

The algorithm for verifying the history of a coin is given in Figure \ref{fig:transfer_coin}

\begin{figure}
\begin{minipage}{\columnwidth}
\begin{framed}
\centering { \bf Protocol for proving and verifying a coin's history } 

The prover sends two lists of \texttt{IncludedTx} elements to the Verifier, \texttt{inclTx} and \texttt{exclTx}. The verifier has access to the Merkle Roots from the deployed Plasma Contract on Ethereum via the variable \texttt{root[blockNumber]} as well as all the committed block numbers for the coin, via the variable \texttt{blocks}.

We consider a VerifyMerkleProof(slot, hash, proof, blockNumber) algorithm which is able to verify the merkle proof for a coin at a certain block number.

\begin{algorithmic}[1]
    \State ${inclTx, exclTx} \gets proof$
    \State // Ensure completion of included and excluded transactions' blocks
    \State Assert $inclTx.keys \cup exclTx.keys = blocks$ 
    \State // Ensure separation between blocks in included and excluded transactions
    \State Assert $inclTx.keys \cap exclTx.keys = \emptyset$
    \State $LastBlock \gets DepositBlock$
    \State $LastOwner \gets DepositOwner$

    \State // Check the deposit transaction
    \If{!VerifyMerkleProof(slot, inclTx[DepositBlock].tx.hash, itx[DepositBlock].proof, itx[DepositBlock].blkNumber)}
        \State \Return false
    \EndIf
    \State delete inclTx[DepositBlock]
    \State // Check that included transactions are correct in the included blocks
    \For {itx in inclTx} // Skip the deposit transaction
    \If{!VerifyMerkleProof(slot, itx.tx.hash, itx.proof, itx.blkNumber)}
        \State \Return false
    \EndIf

    \State // Reject double spends
    \If{$LastBlock \neq itx.tx.parentBlock$}
        \State \Return false
    \EndIf

    \State // Accept spends only with valid signatures
    \State $Sender \gets ecrecover(itx.tx.hash, itx.tx.sig)$
    \If{$Sender \neq LastOwner$}
        \State \Return false
    \EndIf

    \State $LastBlock \gets itx.blkNumber$
    \State $LastOwner \gets itx.tx.newOwner$
    \EndFor

    \State // Check that there are no transactions in the excluded blocks
    \For {itx in exclTx} // itx.tx should be empty
        \If{!VerifyMerkleProof(slot, emptyHash, itx.proof, itx.blkNumber)}
            \State \Return false
        \EndIf
    \EndFor
\end{algorithmic}

\end{framed}
\end{minipage}
\caption{Proving and verifying a coin's history in Plasma Cash}
\label{fig:transfer_coin}
\end{figure}

\subsection{Exiting and Withdrawing a Coin} \label{exiting_withdrawing}
As described in Section \ref{ch2:classic_plasma}, exits are the mechanism by
which a coin can be withdrawn from the plasmachain, and allow it to be
transferred back to its owner's account on the mainchain. 

Starting an exit for a coin requires providing the transaction that gave the
exitor ownership of the coin signed by the previous owner in the coin's history, \textbf{tx}, as well
as a direct ancestor of that transaction (the reason the parent transaction must also be provided is explained in Section 4). Merkle proofs of inclusion need to also be provided for both transactions.

\begin{figure}[H]
	\makebox[\linewidth]{
		\scalebox{0.6}{
		\includegraphics[width=\linewidth]{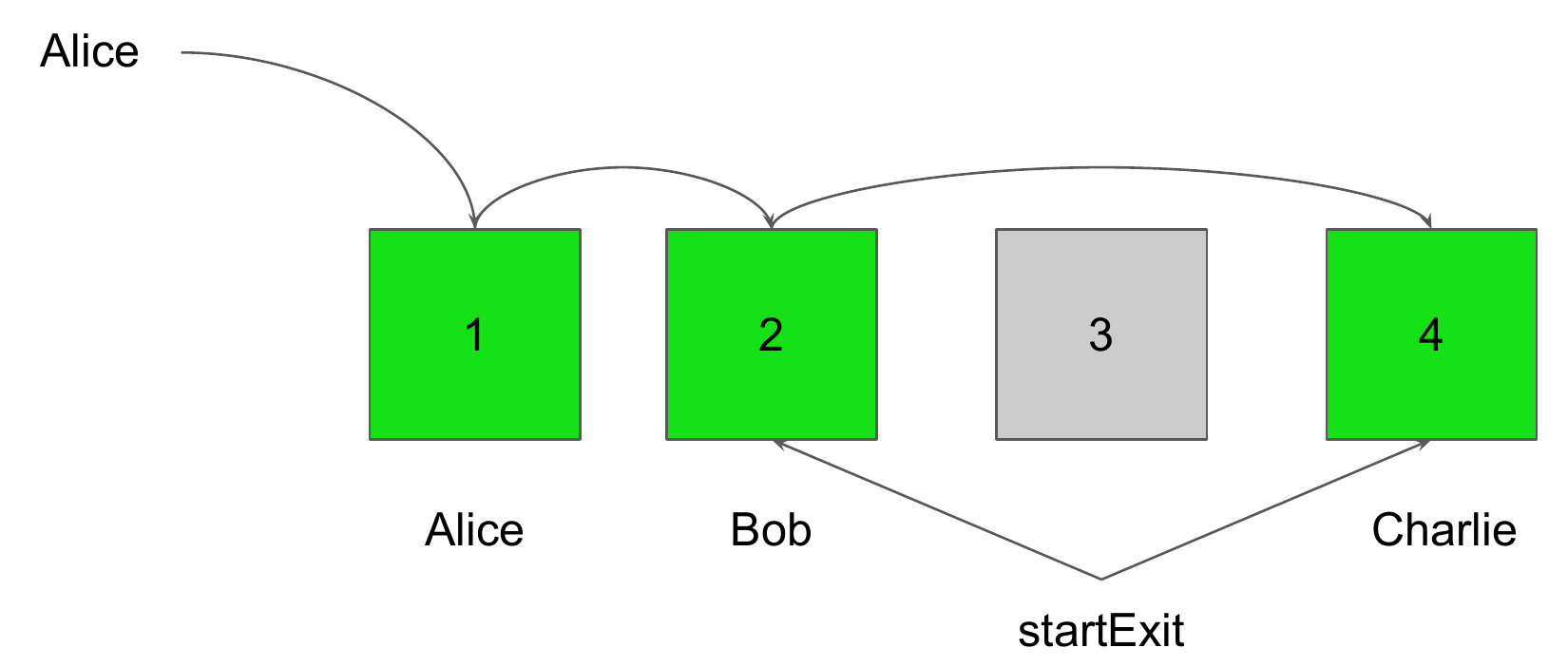}
		}
		}
	\caption{
        Alice deposits a coin from the mainchain to the plasmachain in Block 1. Alice sends the coin to Bob in Block 2. Bob verifies the inclusion of the coin in Block 2. Block 3 gets submitted, without including the coin. Bob sends the coin to Charlie in Block 4. Charlie has to verify the inclusion of the coin in Block 1 and 2, and the non-inclusion of the coin in Block 3. 
        In order for Charlie to exit the coin received by Bob he has to provide
        the signed transaction from Bob as well as a direct ancestor, in this
        case the transaction from Alice to Bob. Charlie also needs to supply
        merkle proofs of inclusion for both of these transactions at their
        respective blocks. 
	}
    \label{fig:exit_lifetime}
\end{figure}

\begin{figure}[H]
	\makebox[\linewidth]{
		\scalebox{0.7}{
		\includegraphics[width=\linewidth]{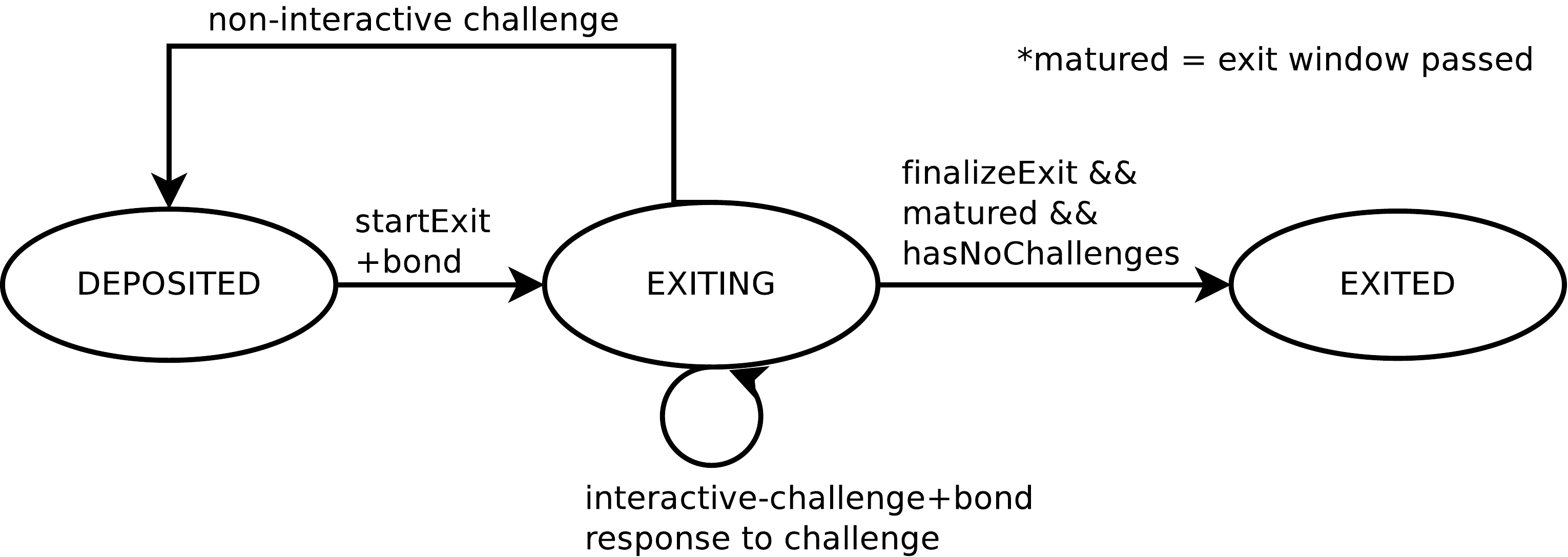}
		}
		}
	\caption{
        The stages of an exit. After a coin transitions to the \texttt{EXITED}
        state, it can be withdrawn to its owner's wallet.
	}
    \label{fig:exit_state_machine}
\end{figure}

A coin can be modelled by a state machine. After starting an exit, the coin transitions to the \texttt{EXITING} 
state. After the challenge (or maturity) period passes, 
the coin's exit can be finalized and it can transition to the \texttt{EXITED} 
state, from which it can be withdrawn to a user's wallet, 
as shown in Figure
\ref{fig:exit_state_machine}. Figure \ref{fig:exit_lifetime} illustrates the 
lifetime of an exit from its initialization to its finalization. 
We further discuss challenges in Section \ref{ch:attacks}.

\begin{figure}[H]
	\makebox[\linewidth]{
		\scalebox{0.8}{
		\includegraphics[width=\linewidth]{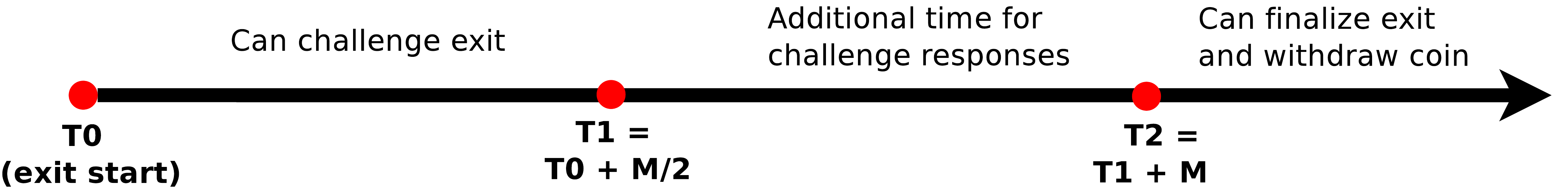}
		}
		}
	\caption{
		During its lifetime, an exit can be challenged during the maturity period. After its maturity period is over, it can be
        finalized and the exiting coin can be withdrawn.
	}
    \label{fig:exit_lifetime}
\end{figure}

\section{Attacks on Plasma Cash} \label{ch:attacks}

In this section we model the types of fraud attempts that users can engage in with or without collusion with the plasmachain's consensus mechanism. We create challenges that guard against malicious behavior, guaranteeing safety as long as the party being attacked logs in\footnote{Once the user logs in, they should challenge any exits that have been initiated for the coins they own} at least once during the dispute period. The challenges described were first introduced in \cite{plasma_cash, plasma_cash_simple_spec} and are further explained in this
paper for clarity.

\subsection{Exit Spent Coin} \label{exit_spent_coin}

\begin{figure}[H]
	\makebox[\linewidth]{
		\scalebox{0.8}{
		\includegraphics[width=\linewidth]{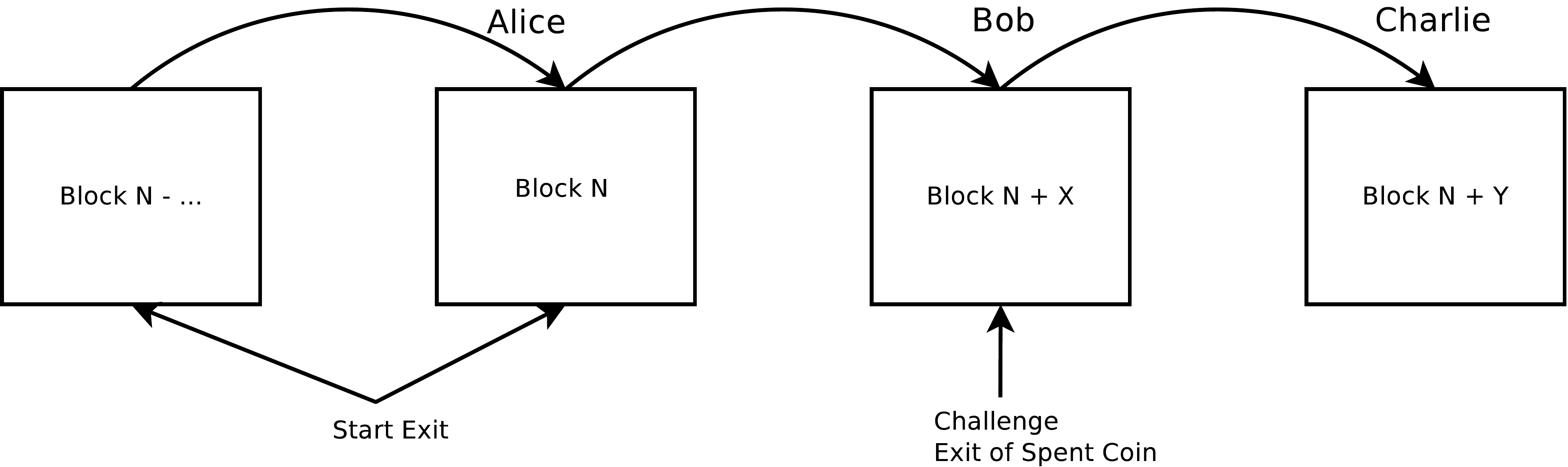}
		}
		}
	\caption{
        Alice attempts to exit the coin that was included in Block $N$. A
        valid challenge provides the inclusion of the transaction at Block
        $N+X$. Challenging with the transaction from Block $N+Y$ is not valid
        since it assumes the validity of its ancestors.
	}
    \label{fig:challenge_after}
\end{figure}

This is an attack which does not require collusion of any of the transacting
parties with the operator. An attacker (Alice) sends a coin to a victim (Bob). 
After both parties verify its inclusion, the attacker immediately attempts to exit the spent coin\footnote{Alice also supplies a security bond as discussed previously, to incentivize users to challenge if her exit is fraudulent, like in this case}. Bob must log in before the end of the challenge period and provide proof of a direct spend of the coin. An example of this is shown in Figure \ref{fig:challenge_after}.

In the reference implementation, this is implemented via
\texttt{challengeAfter}.

\subsection{Exit of a Double Spend}

This attack requires that the operator allows the inclusion of double spend
transactions (ie. two transactions involving the same coin and ancestor
transactions but different new owners included in two different blocks). 
In this case, The attacker (Alice) sends a coin to the victim (Bob). 
After the inclusion of the transaction to Bob, Alice sends another transaction to her colluding party (Charlie). Note that the operator here should notice that Alice no longer owns the coin and reject the transaction, however we consider that they are also colluding with Alice to steal Bob's coin. From the mainchain contract's point of view, both Bob and Charlie are valid owners of the coin. 

\begin{figure}[H]
	\makebox[\linewidth]{
		\scalebox{0.8}{
		\includegraphics[width=\linewidth]{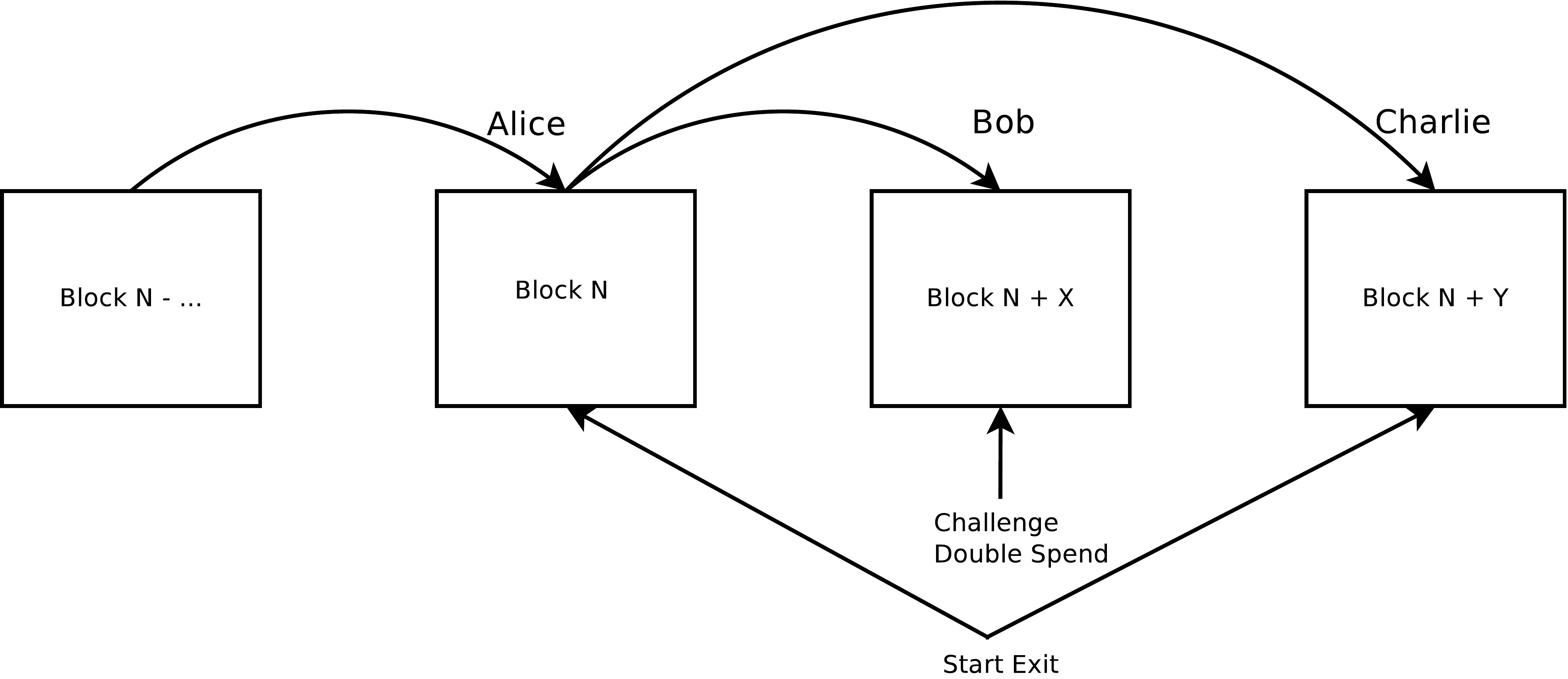}
		}
		}
	\caption{
        Charlie attempts to exit the coin that was included in Block $N+Y$,
        which is a double spend since the coin was already given to Bob at
        block $N+X$. A valid challenge provides the inclusion of the
        transaction at Block $N+X$. 
    }
    \label{fig:challenge_between}
\end{figure}

Charlie can exit by providing the same parent transaction as the one Bob, given that they both received it from the same spend of Alice. A valid challenge involves a transaction between Charlie's exiting and parent block, proving the double spend, as shown in Figure \ref{fig:challenge_between}. Users should additionally check the coin's history for double spends when receiving a coin. In the below example, Charlie can send the coin to another user, the verification of the merkle proofs will pass, however if they find transactions in the coin history with the same parent they should only accept the transaction coming from the earliest owner of the coin, since all the others are double spends.

In the reference implementation, this is implemented via \texttt{challengeBetween}.

\subsection{Exit with Invalid History} \label{exit_with_invalid_history}

This type of attack requires collusion of the transacting parties with the operator to
include transactions which do not have valid ancestors. As an example, consider
the case that the victim (Alice) has a coin that she does not intend to spend. The attacker (Bob) colludes with the operator to create a transaction that
references an invalid transaction from Alice to Bob, effectively giving Bob
ownership of the coin, and sends that transaction to his colluding party (Charlie). After the transaction gets included, the transaction to Charlie can be used 
as a valid ancestor transaction. Taking advantage of that, Charlie sends the coin to
Dylan (another colluding party). From Dylan's perspective, he has a valid transaction signed from the
previous owner of the coin, along with a valid ancestor which is the
transaction from Bob to Charlie. Dylan can now perform a seemingly valid exit.

Note that precisely because of this type of attack, we require that users must validate the full history of a coin they receive in a transaction. If Charlie was not a colluding party and they were a victim as well, he would not accept the transaction because he would have noticed that the coin has an invalid history.

\begin{figure}[H]
	\makebox[\linewidth]{
		\scalebox{0.8}{
		\includegraphics[width=\linewidth]{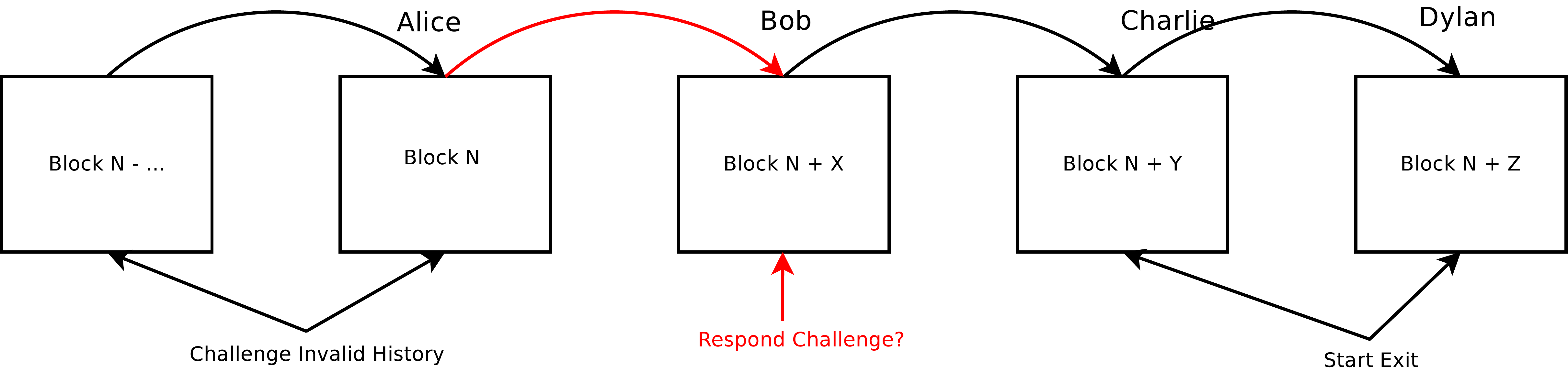}
		}
		}
	\caption{
        Dylan attempts to exit a coin that has invalid history. Challenger
        challenges by claiming that Alice is the last valid owner of the coin. A
        valid response must be a valid transaction that nullifies that claim.
        In this case this does not exist in this case since Bob and the Operator
        colluded to have the invalid transaction included. An Alice-to-Bob
        transaction could be used as a valid response as illustrated by the red
        arrows.
    }
    \label{fig:challenge_before}
\end{figure}

When Dylan initiates the exit, Alice notices that and submits an interactive
challenge, with which she claims that she is the latest valid owner of the coin.
We require this challenge to be interactive and bonded, as it may be the case
that Alice is not the last valid owner of the coin, and a spend of her to another user may exist. 
This leaves a window for a response to her challenge which invalidates it 
(contrary to the previous two challenges which were non-interactive). Multiple of these challenges can be active for an exit. During finalization the coin must have zero pending challenges that were not responded to, in order for the exit to be successful. The full
game is shown in Figure \ref{fig:challenge_before}.

In the reference implementation, this is implemented via \texttt{challengeBefore} and 
\texttt{respondChallengeBefore}.

\subsection{Withhold and Challenge Exit of Spent Coin (Griefing)}

Griefing is a special type of attack which exploits inefficiencies of
the protocol to `bully' the transacting parties and either block them from some action
(disallowing the exit of a coin for example) or forcing them to forfeit some
constant amount of on-chain collateral when they attempt to settle a
transaction on-chain. 

In this attack, consider the scenario where Alice wants to send Bob a coin in
exchange for a product. Bob will only send the product after he has validated
the coin's history and after he has verified that the payment's transaction was included in a block.
This requires that the operator makes the witness data for the transaction's inclusion available, so that both parties can validate the transaction's inclusion.
Instead, the operator publishes a block including the transaction, however they withhold the witness data.

\begin{figure}[H]
	\makebox[\linewidth]{
		\scalebox{0.6}{
		\includegraphics[width=\linewidth]{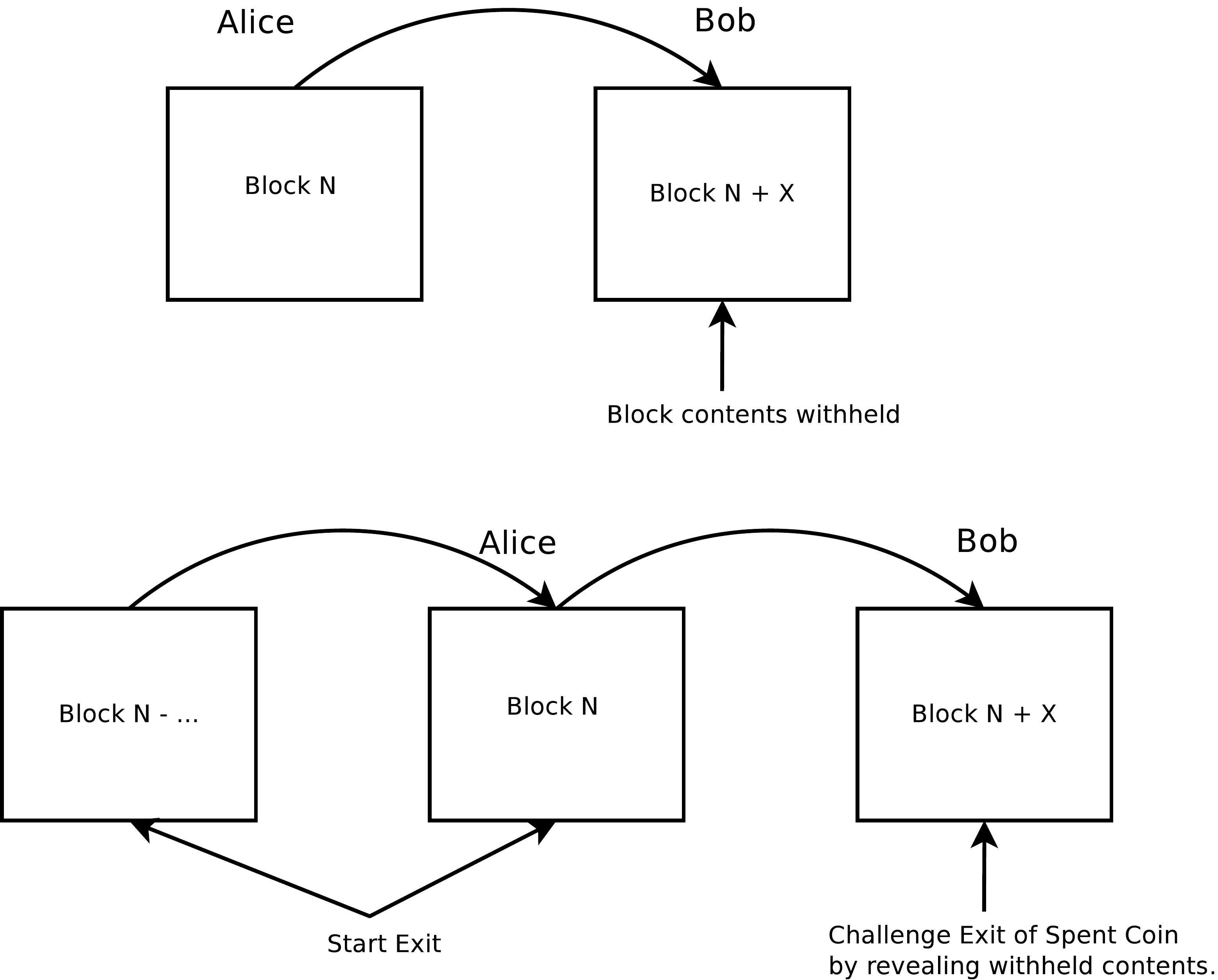}
		}
		}
	\caption{
        The operator steals the bond but the transaction was revealed.
    }
    \label{fig:griefing_withhold}
\end{figure}

As a result, neither Alice nor Bob can know if the transaction was actually included. At this point, Alice must assume that the operator is malicious and initiate an exit for the coin (she can pay Bob through an on-chain transaction). The operator can now challenge the exit with a \textit{Exit Spent Coin}
challenge, as discussed in Section \ref{exit_spent_coin}. During this process,
they are required to reveal witness data (which they were previously
withholding). This provides enough information for both parties to know that the transaction
was successfully included in a block. However, in this process Alice incurred a constant griefing vector by losing the collateral she put up when initiating the exit as shown in Figure \ref{fig:griefing_withhold}.

This can be mitigated by constructing a different type of exit that forces the
settlement of a withheld transaction to a receiving party other than the exitor. The
coined term for this type of exit has been `limbo exit' and the transaction
being withheld is an `in-flight' transaction, given that no party other than
the operator can attest that the transaction has settled or not (and thus is in
`limbo' or still `in-flight') \cite{limbo_exit}.

\section{Future Work}
In this section we go summarize the existing proposals for improving the Plasma
Cash protocol. In-depth analysis of each technique is not in scope and the
reader is encouraged to refer to the citations for further insight.

\subsection{Arbitrary Denomination Payments}

The protocol for Plasma Cash is simplified compared to other Plasma versions
due to the usage of non fungible tokens. This has the disadvantage that for financial use cases, such as buying a product, a user has to provide a coin with the exact product
value, or provide a coin with higher value and expect another coin as change,
similar to how physical cash is used in real life. 

\paragraph{Splits and Merges} Proposals to allow coins to be split and merged are made in  \cite{xuanji_split_merge, dan_split_merge}. Splitting a token involves creating new tokens that
represent a previously unified token. 
Merging involves creating a new token out of a list of separate unspent tokens.

\paragraph{Plasma Debit} A protocol that allows arbitrary denomination payments of a coin is Plasma
Debit which acts as an extension to Plasma Cash \cite{plasma_debit}. In Plasma
Debit, instead of having a set denomination, each coin has a capacity value $v$
and a varying value $a$\footnote{Each coin can be thought of as an account with maximum capacity of $v$}. The following conditions apply:

\begin{enumerate}
    \item after a coin is deposited $v = a$.
    \item $a \leq v$
    \item $a$ of the coin's denomination is owned by the coin's owner
    \item $v - a$ of the coin's denomination is owned by the operator
\end{enumerate}

Each coin in Plasma Debit can be considered as a bidirectional payment channel
between the Plasma Operator and the coin's owner. Arbitrary payments between two
users can be initiated by reducing the balance of the sender's coin by an
amount, and by atomically\footnote{Either all state transitions execute simultaneously, or none} increasing the balance of one of the receiver's coins
by the same amount. The primary limitations in this construction is that the
receiver must have a coin that is undercollateralized (e.g to receive X currency units, it must be the case that $X + a \leq v$). 

Regarding the creation of undercollateralized coins, the protocol can be
extended to allow the operator to deposit collateral, in order to
generate coins with $a = 0$ on deposit, or increase the $v$ while keeping $a$ the same of an already existing coin. Markets can be expected to form where undercollateralized coins with
larger $v$ have more value than coins with smaller $v$, similarly to how
payment channels with more capacity are more useful due to their increased
routing capabilities.

Plasma Debit is a useful construct not only because it allows for arbitrary
payments in Plasma Cash, but because it allows for behavior that can be
characterized as transferrable-payment-channels. While traditional payment
channels between two parties are non transferrable and require knowing all
transacting parties beforehand, transferring a Plasma Debit coin from Alice to Bob is
effectively the transfer of a payment channel between Alice and the operator to
Bob.

\paragraph{Plasma Defragmentation} When depositing a coin, users get multiple coins of very small equal value totalling the value of the deposited coin. Making a payment involves sending multiple of these small denomination coins to the receiver. By exiting a subtree of coins with the same owner, a user can efficiently withdraw and transact in arbitrary denominations. Users that utilize this technique eventually end up with various coins fragmented coins that are not efficiently exitable by a subtree. Defragmentation solves that by rearranging coins between the transacting parties so that users maintain lists with coins that have the most consecutive ids possible, allowing efficient subtree multi-exits.

\subsection{Reduce user data checking}

As discussed in Section \ref{verify_coin_history}, a receiver of a coin has to
verify its history in order to make sure that the coin they are receiving is
valid. Otherwise, they may be receiving a coin with invalid history, which
if exited would be susceptible to \textit{Challenge with Invalid History} from Section
\ref{exit_with_invalid_history}. A proof of inclusion or non-inclusion for a
coin in a chain with $2^{16}$ coins is 512 bytes, per block. Assuming a Plasma
Block frequency of 15 seconds\footnote{The Ethereum mainnet block time}, this
means that after a year, a sender would be required to transfer more than 1 GB of
data of proofs to the receiver, so that the receiver can validate the coin's
history. This is expensive both in terms of bandwith and storage.

\paragraph{Checkpointing of coin history} A solution to this problem is checkpointing the history of coins \cite{plasma_xt}. A plasmachain can
be expected to have a large number of coins, and as a result a mechanism for
efficient mass checkpointing of coins needs to be constructed. After a coin
gets checkpointed, proving the validity of a coin's history requires checking the
history of the coin from its latest checkpoint until the current block, which reduces the amount of merkle proofs that need to be validated.

\paragraph{Compressing non-inclusion proofs} The largest part of data checking in Plasma Cash stems from the verification of non-inclusion proofs. ZK-SNARKs can be used to create a succinct constant-size proof of non-inclusion for multiple blocks for a coin which can be transmitted at low bandwith cost, and can be verified by the receiver in constant-time. Alternatively, a bloom filter hash can be published alongside the merkle root at each block by the operator, succinctly committing to the coin ids spent in each block, allowing users to verify the non-inclusion of a coin in blocks. This approach requires data availability from the plasma operator so that users can retrieve the raw bloom filter from its published hash, as well as revealing the bloom filter on-chain during an exit, which is expensive in storage terms. Finally, after an initialization phase involving a trapdoor parameter, an RSA accumulator can be published alongside the merkle root at each block by the operator, which allows batch proofs of non-inclusion of a coin, effectively reducing the amount of proofs required for the validation of a coin to two (one for its inclusion in a block, and an aggregate proof for all the non-inclusions) \cite{rsa_accum}.

\subsection{Faster and Inexpensive Withdrawals}

\paragraph{Tokenized Exits} A user needs to wait a predefined challenge period when initiating an exit, in order to allow other users to verify that the exit is valid. User experience can be improved by allowing the exitor to instantly redeem the value of their exit by tokenizing it as a NFT and selling it in an open marketplace. The buyer of the tokenized exit will wait the challenge period instead of the exitor \cite{fast_withdrawals}. The tokenized exit's buyout price is equal to the value of the coin that is under the exit. The tokenized exit can be expected to be sold at a discount, depending on the exitor's time preference.

\paragraph{Optimistic Exits} By assuming that a user is honest, the merkle proofs and the raw transactions can be omitted during the initiation of an exit. This reduces the exit's required transaction fees. An additional non-interactive challenge must be provided which reveals the previously omitted exit parameters and if the challenge is successful cancels the exit \cite{optimistic}.

\section{Conclusion}
In this paper we presented Plasma Cash, a plasma construction which can be used to construct non-custodial arbitrarily centralized sidechains which maintain safety, contrary to two-way pegged sidechains which do not allow the safe withdrawal of funds in case the sidechain consensus is compromised. Plasma Cash can be used to offload computation from the mainchain and improve scalability, which is a pressing issue for decentralized permissionless blockchains. We presented the required mechanisms as well as examples of how users interact with the system in order to deposit, transact and withdraw their funds. A reference implementation which is used in production is provided. Further improvements to the protocol are described, with respect to allowing arbitrary denomination payments, improving user experience, and making light clients more efficient.

\bibliographystyle{unsrt}
\bibliography{references}

\begin{thebibliography}{10}

\bibitem{sidechains}
Adam Back, Matt Corallo, Luke Dashjr, Mark Friedenbach, Gregory Maxwell, Andrew
  Miller, Andrew Poelstra, Jorge Timón, and Pieter Wuille.
\newblock {Enabling Blockchain Innovations with Pegged Sidechains}.
\newblock \url{https://blockstream.com/sidechains.pdf}, Oct 2014.

\bibitem{plasma_cash}
Vitalik Buterin.
\newblock {Plasma Cash: Plasma with much less per-user data checking}.
\newblock
  \url{https://ethresear.ch/t/plasma-cash-plasma-with-much-less-per-user-data-checking/1298}.

\bibitem{plasma}
Joseph Poon and Vitalik Buterin.
\newblock {Plasma: Scalable Autonomous Smart Contracts}.
\newblock \url{https://plasma.io/}.

\bibitem{implementers_call}
{Plasma Implementers Call}.
\newblock \url{https://www.youtube.com/channel/UCG2MeKuKDJRK4gFNk-dQuZQ}.

\bibitem{nipopows}
Kiayias Aggelos, Andrew Miller, and Dionysis Zindros.
\newblock {Non-Interactive Proofs of Proof-of-Work}.
\newblock \url{https://eprint.iacr.org/2017/963.pdf}, May 2018.

\bibitem{battleship}
Patrick McCorry, Chris Buckland, Surya Bakshi, Karl Wust, and Andrew Miller.
\newblock {You sank my battleship! A case study to evaluate state channels as a
  scaling solution for cryptocurrencies}.
\newblock \url{https://nms.kcl.ac.uk/patrick.mccorry/battleship.pdf}, Oct 2018.

\bibitem{counterfactual}
Jeff Coleman, Liam Horne, and Li~Xuanji.
\newblock Counterfactual: Generalized state channels.
\newblock \url{https://l4.ventures/papers/statechannels.pdf}.

\bibitem{sparse_merkle_trees}
Rasmus Dahlberg, Tobias Pulls, and Roel Peeters.
\newblock Efficient sparse merkle trees - caching strategies and secure
  (non-)membership proofs.
\newblock {\em IACR Cryptology ePrint Archive}, 2016:683, 2016.

\bibitem{smt}
{What’s a Sparse Merkle Tree?}
\newblock
  \url{https://medium.com/@kelvinfichter/whats-a-sparse-merkle-tree-acda70aeb837}.

\bibitem{smt_compact_proofs}
Plasma cash with sparse merkle trees, bloom filters, and probabilistic
  transfers.
\newblock
  \url{https://ethresear.ch/t/plasma-cash-with-sparse-merkle-trees-bloom-filters-and-probabilistic-transfers/2006}.

\bibitem{plasma_cash_simple_spec}
Plasma cash simple spec.
\newblock \url{https://karl.tech/plasma-cash-simple-spec/}.

\bibitem{limbo_exit}
{Li, Xuanji}.
\newblock {Limbo Exits and Challenging Fraudulent Exits}.
\newblock
  \url{https://ethresear.ch/t/limbo-exits-and-challenging-fraudulent-exits/2015}.

\bibitem{xuanji_split_merge}
One proposal for plasma cash with coin splitting and merging.
\newblock
  \url{https://ethresear.ch/t/one-proposal-for-plasma-cash-with-coin-splitting-and-merging/1447}.

\bibitem{dan_split_merge}
Dan robinson on splitting and merging in plasma cash discussion.
\newblock
  \url{https://ethresear.ch/t/plasma-cash-plasma-with-much-less-per-user-data-checking/1298/53}.

\bibitem{plasma_debit}
Plasma debit: Arbitrary-denomination payments in plasma cash.
\newblock
  \url{https://ethresear.ch/t/plasma-debit-arbitrary-denomination-payments-in-plasma-cash/2198}.

\bibitem{plasma_xt}
Plasma xt: Plasma cash with much less per-user data checking.
\newblock
  \url{https://ethresear.ch/t/plasma-xt-plasma-cash-with-much-less-per-user-data-checking/1926/}.

\bibitem{rsa_accum}
{Buterin, Vitalik}.
\newblock {RSA Accumulators for Plasma Cash history reduction}.
\newblock
  \url{https://ethresear.ch/t/rsa-accumulators-for-plasma-cash-history-reduction/3739/}.

\bibitem{fast_withdrawals}
Kelvin Fichter.
\newblock {Simple Fast Withdrawals}.
\newblock \url{https://ethresear.ch/t/simple-fast-withdrawals/2128/1}.

\bibitem{optimistic}
{Buterin, Vitalik}.
\newblock {Optimistic cheap multi-exit for Plasma (Cash or MVP)}.
\newblock
  \url{https://ethresear.ch/t/optimistic-cheap-multi-exit-for-plasma-cash-or-mvp/1893}.

\end{thebibliography}

\end{document}